\documentclass[aps,pra,twocolumn]{revtex4}
\usepackage{graphicx}
\begin{document}
\title{Collective Atomic Recoil Lasing Including Friction and Diffusion Effects}
\author{G.R.M.\ Robb$^{a}$, N.\ Piovella$^{b}$,  A. Ferraro$^{b}$, and R. Bonifacio$^{b}$}
\affiliation{
$^{a}$Department of Physics, University of Strathclyde, Glasgow, G4 0NG, Scotland.\\
$^{b}$Dipartimento di Fisica, Universit\`a Degli Studi di Milano and INFM,
Via Celoria 16, I-20133 Milano, Italy.}
\author{Ph.W. Courteille and C. Zimmermann}
\affiliation{Physikalisches Institut, Eberhard-Karls-Universit\"at T\"ubingen,
\\Auf der Morgenstelle 14, D-72076 T\"ubingen, Germany}
\date{\today}
\affiliation{}
\begin{abstract}
We extend the Collective Atomic Recoil Lasing (CARL) model including the effects of friction and diffusion forces
acting on the atoms due to the presence of optical molasses fields. The results from this model are consistent with 
those from a recent experiment by Kruse et al. [Phys. Rev. Lett. {\bf 91}, 183601 (2003)]. In particular, we obtain a threshold 
condition above which collective backscattering occurs. Using a nonlinear analysis we show that the backscattered field 
and the bunching evolve to a steady-state, in contrast to the non-stationary behaviour of the standard CARL 
model. For a proper choice of the parameters, this steady-state can be superfluorescent.
\end{abstract}

\maketitle
The mechanical effect of light on atoms has now been the subject of intense theoretical and 
experimental research efforts for several decades. However, the fact that the collective atomic centre-of-mass motion 
of atoms can strongly influence the evolution of optical fields has only received attention relatively 
recently \cite{CARL:1,CARL:2,CARL:3,Narducci}. Recent experimental studies involving large numbers of cold atoms in 
high-quality cavities \cite{Kruse,Hemmerich,Vuletic} represent an important advance in this field, allowing detailed 
experimental studies of collective atom-light interaction dynamics. During these interactions both the mechanical 
effect of the cavity modes on the atomic motion and the driving of the cavity modes by the dynamic spatial distribution 
of atoms in the cavity must be described self-consistently and cannot be considered independently.

A recent experiment by Kruse et al. \cite{Kruse} represents the first unambiguous realization of the CARL model 
originally proposed by Bonifacio and coworkers \cite{CARL:1}, which describes collective backscattering of an optical 
pump field by a sample of cold atoms.
Previous experiments on CARL have been performed in hot atomic vapours \cite{CARL:exp1,CARL:exp2} where however
the gain of the backward field can not be unambiguously attributed to atomic recoil.
 
Here we extend the previous theoretical work on CARL, including
the effects of friction and diffusion forces acting on the atoms due to the presence of optical molasses fields. 
We describe the system using a set of coupled Maxwell-Fokker-Planck equations. 
A linear stability analysis reveals that there is a threshold condition for the pump power 
above which collective backscattering occurs.
Preliminary experimental results confirm this prediction \cite{Zimmermann}. We further show
that the backscattered field and atomic density modulation amplitude or `bunching' 
evolve to a steady-state, in contrast to the non-stationary behaviour observed using the standard CARL model. 
Our model describes the main features of the experimental results of \cite{Kruse}.

In addition to the so-called `good-cavity' regime in which the experiment of \cite{Kruse} operates, we also examine 
the behaviour of the system in the `bad-cavity' regime.  We show that in this regime the atoms emit in a
superfluorescent way \cite{SF}, with scattered intensity $\propto N^2$, where $N$ is the number of atoms.
This is a unique example of a steady-state superfluorescence, i.e. superradiance from an incoherently prepared atomic 
system.

Our system consists of an ensemble of atoms backscattering a far-detuned pump field into a 
counterpropagating mode of a ring cavity, described by a set of equations derived by Bonifacio and 
coworkers \cite{CARL:1,CARL:2,CARL:3}. The equations are here generalized to include a friction force, 
$- \bar\gamma\bar p$, and a stochastic force, $F(\bar t)$, 
due to the presence of optical molasses:
\begin{eqnarray}
\frac{d \theta}{d \bar t} &=&\bar p \label{dtheta_dt2} \\
\frac{d\bar p}{d \bar t} &=& - \left( A e^{i \theta} + c.c. \right) - \bar\gamma\bar p + F(\bar t) \label{dp_dt2} \\
\frac{d A}{d \bar t} &=& \left \langle e^{-i \theta} \right \rangle - K A \label{dA_dt2}.
\end{eqnarray}
where $\bar t=\omega_r\rho t$, $\theta=2kz$, $\bar p=2kv_z/\omega_r\rho$, and 
$A=(2\epsilon_0V/N\hbar\omega\rho)^{1/2}E$ are universally scaled time, position, momentum and scattered 
electric field variables, respectively, normalized to the CARL bandwidth $\omega_r\rho$ where 
$\omega_r=2\hbar k^2/m$ is the recoil frequency,
$\rho=(\Omega_0/2\Delta)^{2/3}(\omega d^2 N/2V\hbar\epsilon_0\omega_r^2)^{1/3}\propto (P_0 N/\Delta^2)^{1/3}$,
$\Omega_0$ is the Rabi frequency of the pump field with frequency $\omega=ck$, detuned from the atomic resonance by 
$\Delta=\omega-\omega_0$, $d$ is the electric dipole moment of the atom, $P_0$ is the intracavity pump power and
$V$ is the cavity mode volume.
$K=\kappa_c/\omega_r\rho$ represents scaled cavity losses and $\bar\gamma=\gamma_f/\omega_r\rho$ is the scaled damping
coefficient to account for molasses friction.
With the exception of the stochastic force, $F(\bar t)$, these equations are similar to the ones used by 
Bonifacio and Verkerk \cite{BV} to describe CARL including the effect of collisions. 

Here, the stochastic force, $F(\bar t)$, causes diffusion of the atomic momenta i.e. heating. 
We assume white noise, i.e.
$\langle F(\bar t)F(\bar t')\rangle =2D_p\delta(\bar t-\bar t')$,
where $D_p=\bar\gamma\sigma^2$ is the momentum diffusion coefficient and 
$\sigma$ is the momentum spread in units of $\omega_r\rho$, corresponding to the Doppler temperature $T$ of the 
atoms in the molasses fields.

In the limit of strong viscous damping, it is possible to adiabatically eliminate the atomic momentum for each atom 
by setting $\frac{d\bar p}{d \bar t} = 0$ in Eq.~(\ref{dp_dt2}), so that
\begin{equation}
\label{elimination}
\bar p = -\frac{1}{\bar\gamma} \left( A e^{i \theta} + c.c. \right) + \frac{F(\bar t)}{\bar\gamma}.
\end{equation}
A simple steady-state analytical solution can be derived from Eqs.(\ref{dA_dt2}) and (\ref{elimination}) in absence of 
the stochastic force $F$, i.e. neglecting diffusion, and for the case of perfect bunching. In fact, assuming
 $\langle e^{-i\theta}\rangle\approx e^{-i\langle\theta\rangle}$ in Eqs.(\ref{dA_dt2}) and (\ref{elimination}),
 one obtains \cite{Kruse}
\begin {equation}
a=\frac{e^{-i\langle\theta\rangle}}{\kappa-i\langle p\rangle}\qquad\qquad
\langle p\rangle=-\frac{2\kappa}{\kappa^2+\langle p\rangle^2},
\label{perfect}
\end{equation}
where we have normalized all the variables in order to reduce the number of free parameters, defining 
$a=A/\sqrt{\bar\gamma}$, $p=\sqrt{\bar\gamma}\bar p=d\theta/d\tau$, $\tau=\bar t/\sqrt{\bar\gamma}$ and
$\kappa = \sqrt{\bar\gamma}K$. In the following we will use these scaled variables and parameters.

Depending on the value of $\kappa$, it is possible to define two different steady-state regimes for CARL:
when $\kappa\ll 1$ ('good-cavity' limit), we obtain
$|\langle p\rangle|\approx (2\kappa)^{1/3}\gg \kappa$ and $|a|^2\approx (2\kappa)^{-2/3}$. 
Since $\kappa\propto\kappa_c(\gamma_f\Delta^2/NP_0)^{1/2}$ and
$|a|^2\propto P_s/N$, where $P_s$ is the scattered light power, then $P_s\propto N^{4/3}$
as in the usual CARL \cite{CARL:1,CARL:2,CARL:3}. Conversely,  for $\kappa\gg 1$ ('bad-cavity' limit) we obtain
$|\langle p\rangle|\ll \kappa$ and $|a|^2\approx 1/\kappa^2$, so that the scattered  power $P_s$ is 
proportional to $N^2$, i.e. {\it superfluorescent}.

Diffusion can be described, in the adiabatic limit of Eq.(\ref{elimination}), writing a Fokker-Planck equation for the 
distribution function $P (\theta,\tau)$ \cite{FP1,FP2}. Together with 
Eq. (\ref{dA_dt2}), our model becomes:
\begin{eqnarray}
\label{dB_dt}
\frac{\partial P}{\partial\tau} &=& 
\frac{\partial}{\partial \theta}\left [\left( a e^{i \theta} + c.c. \right)   P\right] 
+D \frac{\partial^2 P}{\partial \theta^2}  \\
\label{FP-wave}
\frac{d a}{d\tau} &=& \int_{0}^{2 \pi} P(\theta,\tau)e^{-i \theta} \;d \theta - \kappa a,
\end{eqnarray}
where $D = \sqrt{\bar\gamma}D_\theta=\sigma^2/\sqrt{\bar\gamma}\propto T(\Delta^2/\gamma_f N P_0)^{1/2}$ and 
$D_\theta=\sigma^2/\bar\gamma$ is the space diffusion coefficient.

The distribution function is normalised such that $\int_{0}^{2 \pi} P(\theta,\tau) \;d \theta = 1$ and is
periodic in $\theta$ with period $2 \pi$. Hence, Eq.~(\ref{dB_dt}) and (\ref{FP-wave}) can be written in terms of the 
spatial harmonics of $P(\theta,\tau)$ i.e.
$P(\theta, \tau) = (1/2\pi)\sum_{n=-\infty}^\infty B_n (\tau) e^{i n \theta}$,
where $B_n(\tau)=\int_{0}^{2 \pi} P(\theta,\tau)e^{-in\theta} \;d \theta$, so that
 \begin{eqnarray}
\label{Pk_eqn}
\frac{d B_n}{d\tau} &=& in\left( a B_{n-1} + a^* B_{n+1} \right) - n^2 D B_n \\
\label{A_eqn}
\frac{d a}{d \tau} &=& B_1 - \kappa a.
\end{eqnarray}
We note that $B_{-n} = B_n^*$ and $B_0 = 1$. In particular $b=|B_1|=|\langle e^{-i \theta}\rangle|$ is the bunching 
factor, describing the ampitude of the density grating. Moreover, from Eq.(\ref{elimination}) it follows that the 
average momentum is $\langle p\rangle=-2{\rm Re}(aB_1^*)$.

In the following, we will show that diffusion is responsible for the existence of a definite threshold for the 
steady-state solution, for which in general the bunching factor $b$ is less than one.

The system of equations (\ref{Pk_eqn}) and (\ref{A_eqn}) have a steady-state solution with
$a^{(0)} = 0$ and $B_n^{(0)}=0$ for $n\neq 0$. If we linearise Eq.~(\ref{Pk_eqn}) and (\ref{A_eqn}) around
this steady-state, we find that fluctuations of $a$ and $B_n$ grow exponentially 
$\propto \exp \left( \lambda \tau \right)$  if the dispersion relation
\begin{equation}
\label{dispersion}
\left ( \lambda + \kappa \right) \left( \lambda + D \right) =i
\end{equation}
has roots with positive real parts. The real and imaginary parts of the unstable mode of Eq.(\ref{dispersion}) gives
the gain $G$ and the frequency shift $\Delta\omega$ in the exponential regime, that, in units of the cavity
bandwidth $\kappa_c$, are: 
\begin{eqnarray}
\frac{G}{\kappa_c} &=&\frac{{\rm Re}\lambda}{\kappa}=
\frac{1}{\kappa} \left[\sqrt{\frac{C^2+\sqrt{1+C^4}}{2}} - \frac{\kappa+D}{2} \right]\label{gain}\\
\frac{\Delta \omega}{\kappa_c} &=&\frac{{\rm Im}\lambda}{\kappa}=\frac{1}{\sqrt{2}\kappa} 
\left[\frac {1}{\sqrt{\sqrt{1+C^4}+C^2}} \right] \label{freqshift} 
\end{eqnarray}
where $C =(\kappa - D)/2$.
The threshold condition for instability ($G\ge 0$) gives rise to the following relation between $\kappa$ and $D$:
\begin{equation}
\label{threshold}
\kappa D \left(D + \kappa \right)^2 \leq  1.
\end{equation}

\begin{figure}[h]
\includegraphics[width=0.3\textwidth]{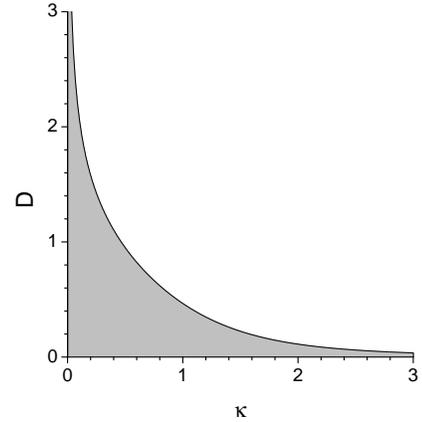}
\caption{Region of instability as a function of $\kappa$ and $D$ (gray area).}
\label{linear}
\end{figure}

Figure~\ref{linear} shows the regions of parameter space ($\kappa, D$) for which Eq. (\ref{threshold}) predicts 
unstable growth of the probe field and the atomic density modulation.

From the experiment of Kruse et al. \cite{Kruse}, the minimum value of the frequency shift measured was
$\Delta \omega / \kappa_c \approx 4.6$. Assuming that it corresponds to the threshold value of the frequency shift,
then, using Eq.(\ref{threshold}) in Eq.(\ref{freqshift}) we calculate
that the values of $\kappa$ and $D$ at threshold are $\kappa\approx 0.1$ and $D\approx 2.1$. 
The experiment of Ref.\cite{Kruse} can therefore be described using the `good cavity' limit where $\kappa \ll 1$. 
In this limit, the threshold condition Eq.(\ref{threshold}) 
becomes simply $D \leq \kappa^{-1/3}$, and the frequency shift at threshold is approximately
$\Delta\omega/\kappa_c\approx \kappa^{-2/3}\approx D^2$. 
Using the definition of $\kappa$, the threshold value $\kappa=0.1$ can be 
expressed as a threshold condition for the pump power. Assuming the same parameters as in \cite{Kruse}, i.e
$\kappa_c=(2\pi)22$kHz, $\gamma_f=9\kappa_c$ and $T\approx 150\mu K$, we obtain a CARL parameter $\rho=14.6$
at threshold, corresponding to an intracavity pump power of $P_0 \approx 3 \mbox{W}$, in good agreement with the 
experimental value.
The experimentally accessible parameters are the pump power and the CARL frequency at threshold.
The linear analysis can provide us with important information on the scaling behaviour 
of the threshold input pump power and the frequency shift at threshold. 
Using the condition $D\le\kappa^{-1/3}$ and the scaling of $D$ and $\kappa$ given above, one obtains that the
intracavity pump power scales as
\begin{equation}
\label{Pscale}
P_0 \propto \frac{T^{3/2} \Delta^2 \kappa_c^{1/2}}{N \gamma_f^{1/2}}.
\end{equation}
In the same 'good-cavity' limit, from Eq.(\ref{freqshift}),
the frequency shift at threshold can be shown to scale as 
\begin{equation}
\label{freqscale}
\frac{\Delta \omega_{\mbox{th}}}{\kappa_c} \propto \kappa^{-2/3} \propto \sqrt{\frac{T}{\gamma_f\kappa_c}}.
\end{equation}

It is also worthwhile to investigate the scaling behavior of the threshold pump power and the frequency 
shift at threshold in the superfluorescent regime, $\kappa\gg 1$. Although to date there have been no experimental 
studies which operate in this regime, we can use our model to predict the behaviour of such an experiment. It should 
be noted that from an inspection of the region of instability shown in figure~\ref{linear}, in order to operate in the 
superfluorescent regime {\em and} remain above the threshold for instability, it is necessary to increase scaled cavity 
losses $\kappa$ and decrease the scaled diffusion/temperature parameter $D$.
In the bad-cavity limit the threshold condition Eq.(\ref{threshold}) becomes simply $D \leq \kappa^{-3}$, 
and the frequency shift at threshold is approximately $\Delta\omega/\kappa_c\approx \kappa^{-2}\approx D^{2/3}$. 
Consequently the threshold pump power in the superfluorescent regime scales as 
\begin{equation}
\label{Pscale_sr}
P_0 \propto \frac{T^{1/2} \Delta^2 \kappa_c^{1/2} \gamma_f^{1/2}}{N}.
\end{equation}
Note that the dependence of threshold pump power on temperature is now $\propto T^{1/2}$ in the bad cavity limit, 
as opposed to $\propto T^{3/2}$ in the good cavity limit. The dependence on $N$ and $\Delta$ is the same as that 
for the good cavity limit.
The frequency shift at threshold in the bad cavity limit can be shown to scale as 
\begin{equation}
\label{freqscale2}
\frac{\Delta \omega_{\mbox{th}}}{\kappa_c} \propto \frac{1}{\kappa^2} \propto \sqrt{\frac{T}{\gamma_f\kappa_c}}
\end{equation}
so the dependence of the frequency shift at threshold on temperature, friction and cavity losses will be the same in 
both the good and bad cavity limits.

\begin{figure}[h]
\includegraphics[width=8cm,height=7cm]{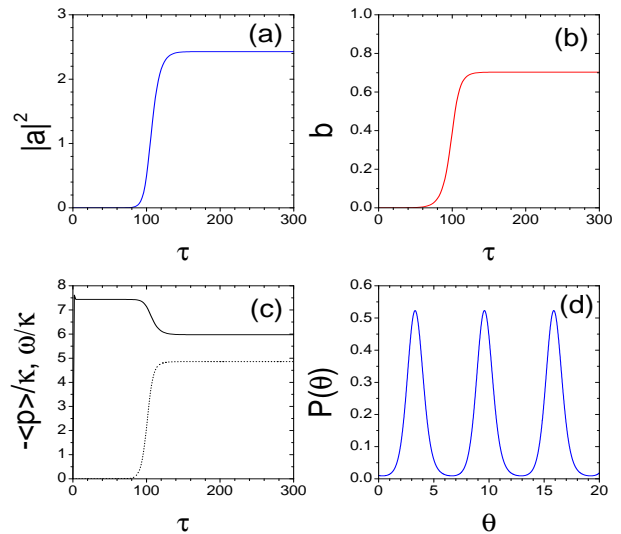}
\caption{Graph of $|a|^2$, (a), $b$, (b), $\omega/\kappa$ (continuous line) and $-\langle p\rangle/\kappa$ 
(dotted line), (c), vs. $\tau$ and of the stationary distribution $P(\theta)$ vs. $\theta$, (d).}
\label{time}
\end{figure}

The numerical integration of Eqs. (\ref{Pk_eqn}) and (\ref{A_eqn}) shows that the system evolves toward the stationary 
solution with $a^{(s)}=\alpha e^{i\omega\tau}$ (where $\omega=d\phi/d\tau$ is the scaled frequency shift) and 
$B_n^{(s)}=\beta_n e^{in\omega\tau}$, where $\alpha$ and $\beta_n$ are complex constants. As a consequence,
the density distribution moves at a constant velocity, $P(\theta,\tau)=P^{(s)}(\theta+\omega\tau)$.
We note that at steady-state the average momentum is $\langle p\rangle=-2\kappa b^2/(\kappa^2+\omega^2)$, where the
bunching parameter is $b=|\beta_1|$. Comparing with Eq.(\ref{perfect}), it follows that in the case of perfect 
bunching ($b=1$) $\omega=-\langle p\rangle$.
Figure~\ref{time} shows the evolution of the backscattered scaled intensity, $|a|^2$, (a), and the bunching 
$b$, (b), as a function of scaled time $\tau$ for parameters close to those of the experiment by Kruse et al. 
i.e. $D=1.49$ and $\kappa=0.075$.
The instability was initiated using a seed field $a_0 = 10^{-5}$.

\begin{figure}[h]
\includegraphics[width=8truecm,height=4truecm]{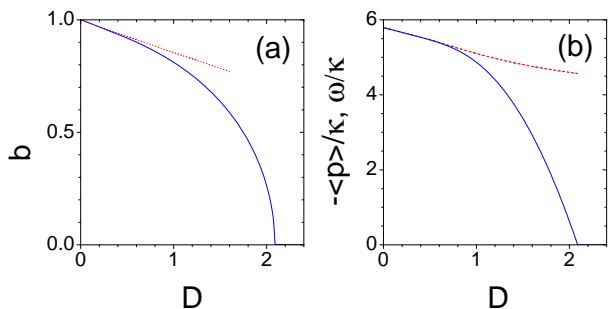}
\caption{(a): Steady-state solution for $b$ (continuous line). 
The dashed line represents the approximated solution (\ref{b_nl}).
(b): Steady-state solution for $\omega/\kappa$ (dashed line) and 
$-\langle p\rangle/\kappa$ (continuous line), as a function of $D$ for $\kappa=0.1$.}
\label{stationary}
\end{figure}

It can be seen that the field intensity and the bunching increase exponentially before relaxing to a steady-state. 
This is the same qualitative behaviour as observed in the experiment \cite{Kruse}. 
Fig.\ref{time} (c) shows also the time evolution of the frequency shift 
$\Delta\omega/\kappa_c=\omega/\kappa$ (continuous line) and of the scaled mean velocity 
$2k\langle v_z\rangle/\kappa_c=\langle p\rangle/\kappa$ (dashed line). 
We observe that the atomic mean velocity at steady-state does not coincide with the velocity of the 
optical standing wave, i.e. $-\langle p\rangle\neq \omega$, as occurs in the case 
of perfect bunching $b=1$. Finally, Fig.\ref{time} (d) shows the stationary distribution $P(\theta)$ vs. $\theta$, 
showing the density grating profile.
The steady-state solution of Eqs. (\ref{Pk_eqn}) and (\ref{A_eqn}) may be obtained solving the following 
recurrence equation for $\beta_n$:
\[(\omega-inD)\beta_n=\alpha\beta_{n-1}+\alpha^*\beta_{n+1},
\]
where $n\neq 0$ and $\alpha=\beta_1/(\kappa+i\omega)$. Calculating $\beta_1$ in terms of a continued fraction and 
iterating numerically the solution in order to find $\omega$, the stationary solution can be obtained exactly.
As an example, Fig.\ref{stationary}a shows the steady-state bunching $b$ (continuous line) and
Fig.\ref{stationary}b shows $-\langle p\rangle/\kappa$ (continuous line)
and $\omega/\kappa$ (dashed line) as a function of $D$ and for $\kappa=0.1$. 
We observe that the bunching parameter $b$ goes to zero at the threshold 
value $D_{th}=2.1$, and the frequency shift at threshold is $\omega=4.6\kappa$, in agreement with the results of 
the linear theory. We note from Fig.\ref{stationary}(b) that $-\langle p\rangle\approx\omega$ only well above threshold,
when $D\ll D_{th}$.

In this limit an approximate solution for the frequency shift $\omega$ and the bunching parameter $b$ can be obtained 
assuming that the steady-state probability distribution $P^{(s)}(\theta+\omega\tau)$ is a Gaussian. In this case it is 
possible to show that $b$ and $\omega$ can be obtained from the solution of the following two coupled equations:
\begin{eqnarray}
\omega &=& \frac{2\kappa b^2}{\kappa^2 +\omega^2}\label{freq_nl} \\
b&=& \exp \left( - \frac{D\sqrt\kappa}{2\sqrt{2\omega - \kappa\omega^2}} \right) \label{b_nl}
\end{eqnarray}
The approximate solution of Eqs.(\ref{freq_nl}) and (\ref{b_nl}) for $b$ (dashed line) is shown by a 
dashed line in fig.\ref{stationary}(a) for comparison with the exact solution. 
We note that the approximate solution is valid only when
$-\langle p\rangle\approx\omega$, as can be seen from fig.\ref{stationary}b.

\begin{figure}[h]
\includegraphics[width=8truecm,height=4truecm,clip=true]{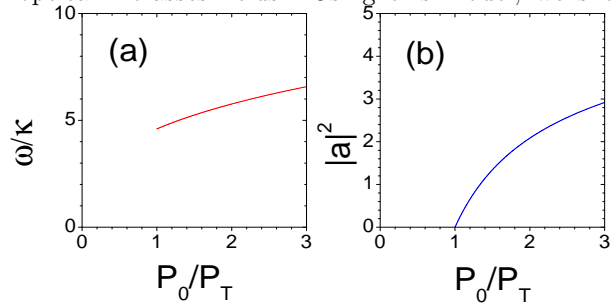}
\caption{Graph of frequency shift $\omega/\kappa$, (a), and scattered power $|a|^2$, (b), 
as a function of $P_0/P_{T}$, for $\kappa_c=(2\pi)22 kHz$, $\gamma_f=9\kappa_c$ and $T=150\mu K$.}
\label{expe}
\end{figure}

In the experiment of ref.\cite{Kruse}, the pump power is ramped up and down, while the frequency shift 
$\omega/\kappa$ is monitored. 
The ramp can be designed in such a way that the pump power crosses the threshold value $P_{T}$, 
so that we expect that the signal of power $P_s$ of the scattered light beam has the 
dependencies shown in Fig.~\ref{expe}. 
This behavior seems to be in agreement with preliminary experimental results, to be discussed elsewhere 
\cite{Zimmermann}.

In conclusion, we have presented a model of the recent experiments by Kruse et al. \cite{Kruse} showing collective 
backscattering 
of an optical pump field by a sample of cold atoms. The model used is an adaptation of the CARL model which includes 
the effects of friction and diffusion forces acting on the atoms due to the presence of optical molasses fields. 
Using this model, we show that the system can be described by a system of coupled Maxwell-Fokker-Planck equations. 
It was shown using a linear stability analysis that there is a threshold condition above which collective 
backscattering occurs. Using a nonlinear analysis it was shown that the backscattered field and atomic density 
modulation amplitude or `bunching' evolves to a steady-state, in contrast to the non-stationary behaviour observed 
using the usual CARL model.
We suggest to test our model experimentally as well in the linear regime by searching for a threshold and 
characterizing its dependence from the coupling parameter and from temperature, as in the nonlinear regime. 
The experiment currently, operates in the so-called `good-cavity' regime \cite{Kruse}. 
We furthermore propose to tune the experiment into the `bad-cavity' regime in which the atoms scatter 
superfluorescently, with scattered intensity $\propto N^2$. This would represent the first \emph{cw} 
superfluorescent system realized so far.

The authors GRMR, NP and RB would like to acknowledge support from the
Royal Society of London via a European Science Exchange Joint Project.

\end{document}